\documentclass{eptcs}
\providecommand{\event}{ADG 2021} 
\usepackage{graphicx}
\usepackage{hyperref} 
\usepackage{amssymb, amsmath}
\usepackage{etoolbox}
\usepackage{listings}
\usepackage{xcolor}
\usepackage{todonotes}
\presetkeys{todonotes}{fancyline}{}
\usepackage{caption}
\usepackage{subcaption}

\usepackage{standalone}
\usepackage{tikz}
\usetikzlibrary{arrows,calc,tikzmark}
\definecolor{tikz-red}{rgb}{0.5019607843137255,0,0}
\definecolor{tikz-darkblue}{rgb}{0,0.2,0.6}
\definecolor{tikz-blue}{rgb}{0.49019607843137253,0.49019607843137253,1}
\definecolor{tikz-gray}{rgb}{0.5,0.5,0.5}

\newcommand{\E}{\mathcal{E}}
\renewcommand{\P}{\mathcal{P}}

\newcommand{\ord}[3]{\ensuremath{[#1\;#2\;#3]}}

\newcommand{\chain}[2][\;\dots\;]{%
 \ensuremath{\left[%
  \def\nextitem{\def\nextitem{#1}}
  \renewcommand*{\do}[1]{\nextitem##1}
  \docsvlist{#2}
 \right]}%
}
\makeatletter
\newcommand{\customlabel}[2]{%
   \protected@write \@auxout {}{\string \newlabel {#1}{{#2}{\thepage}{#2}{#1}{}} }%
   \hypertarget{#1}{#2}%
}
\makeatother

\newtheorem{theorem}{Theorem}
\newtheorem{axiom}{Axiom}
\newtheorem{definition}{Definition}

\def\orcidID#1{\unskip$^{\mbox{\href{https://orcid.org/#1}{\scriptsize{[#1]}} }}$}

\definecolor{isa-rangecomment}{RGB}{139,0,0}
\definecolor{bgr-lightgrey}{gray}{0.98}
\definecolor{isa-kwcolor-2}{RGB}{27,158,119}
\definecolor{isa-kwcolor-3}{RGB}{217,95,2}
\definecolor{isa-kwcolor-4}{RGB}{117,112,179}
\colorlet{isa-proofsnip}{isa-kwcolor-3}
\lstdefinelanguage{Isar}%
{
  morekeywords=[2]{
    _first,
    theory,begin,end,types,datatype,consts,defs,primrec,
    syntax,translations,apply,
    lemma,theorem,corollary,definition,inductive,where,locale,context,
    done,next,proof,and,qed,
    assumes,fixes,shows,case,goal,obtains,defines,
    fixed variables,
    _last
  },
  keywordstyle={[2]\bfseries\color{isa-kwcolor-2}},
  morekeywords=[3]{
  	sorry,oops
  },
  keywordstyle={[3]\bfseries\color{isa-kwcolor-3}},
  morekeywords=[4]{
    then,this,thus,hence,have,assume,obtain,moreover,fix,show,by,using,unfolding,
    ultimately,let,if,for,else,consider
  },
  keywordstyle={[4]\bfseries\color{isa-kwcolor-4}},
  sensitive=true,
  morecomment=[l]{-- },
  morecomment=[s]{(*}{*)},
  literate=
  	{(*...*)}{{\textcolor{isa-proofsnip}{\textbf{...}}}}3
  	{<proof>}{{\textcolor{isa-proofsnip}{\textbf{<proof>}}}}7
    {\\<not>}{{$\neg$}}1
    {\\<times>}{{$×$}}1                 
    {\\<Rightarrow>}{{$\Rightarrow$}}2%
    {\\<equiv>}{{$\equiv$}}1
    {~=}{{$\not=$}}1                         
    {\\<rightleftharpoons>}{{$\rightleftharpoons$}}2
    {\\<exists>}{{$\exists$}}1
    {\\<forall>}{{$\forall$}}1
    {\\<in>}{{$\in$}}1
    {\\<notin>}{{$\notin$}}1
    {\\<union>}{{$\cup$}}1
    {\\<Union>}{{$\bigcup$}}2
    {\\<inter>}{{$\cap$}}1
    {\\<emptyset>}{{$\emptyset$}}1
    {\\<theta>}{{$\theta$}}1
    {\\<Theta>}{{$\Theta$}}1
    {\\<lambda>}{{$\lambda$}}1
    {\\<E>}{{$\E$}}1
    {\\<subseteq>}{{$\subseteq$}}1
    {\\<ge>}{{$\geq$}}1
    {\\<le>}{{$\leq$}}1
    {\\<noteq>}{{$\neq$}}1
    {\\<P>}{{$\P$}}1
    {\\<parallel>}{{$\parallel$}}1
    {\\<Longrightarrow>}{{$\Longrightarrow$}}3
    {\\<longrightarrow>}{{$\longrightarrow$}}3
    {\\<leftrightarrow>}{{$\leftrightarrow$}}2
    {\\<longleftrightarrow>}{{$\longleftrightarrow$}}3
    {\\<lbrakk>}{{$[\![$}}1
    {\\<rbrakk>}{{$]\!]$}}1
    {\\<and>}{{$\land$}}1
    {\\<or>}{{$\lor$}}1
    {\\<bigwedge>}{{$\bigwedge$}}1
    {\\<And>}{{$\bigwedge$}}1
    {\\<triangle>}{{$\triangle$}}1
    {\\<open>}{{$\langle$}}1
    {\\<close>}{{$\rangle$}}1
    {\\<\^cancel>}{{\%}}1
    {\\<\^sub>}{{\_}}1
    {ü}{"u}1 {ä}{"a}1 {ö}{"o}1
    {Ü}{"U}1 {Ä}{"A}1 {Ö}{"O}1 {ß}{{ss}}1
}[keywords,comments,strings]%

\renewcommand{\lstinline}{\oldlstinline[basicstyle=\ttfamily\normalsize]}

\begin{document}

\title{Formalising Geometric Axioms for Minkowski Spacetime 
and Without-Loss-of-Generality Theorems}

\author{Richard Schmoetten\orcidID{0000-0003-1473-071X} 
\institute{Artificial Intelligence and its Applications Institute \\
School of Informatics, University of Edinburgh \\
Edinburgh EH8 9AB, United Kingdom}
\email{richard.schmoetten@ed.ac.uk} 
\and
Jake Palmer\orcidID{0000-0002-4550-0581} 
\institute{Artificial Intelligence and its Applications Institute \\
School of Informatics, University of Edinburgh \\
Edinburgh EH8 9AB, United Kingdom}
\email{jake.palmer@ed.ac.uk}
\and
Jacques Fleuriot\orcidID{0000-0002-6867-9836}
\institute{Artificial Intelligence and its Applications Institute \\
School of Informatics, University of Edinburgh \\
Edinburgh EH8 9AB, United Kingdom}
\email{jdf@ed.ac.uk}
}

\def\titlerunning{Formalising Geometric Axioms for Minkowski Spacetime 
and WLOG Theorems}
\def\authorrunning{Schmoetten, Palmer \& Fleuriot}

\maketitle              

\begin{abstract} 
This contribution reports on the continued formalisation of an axiomatic system for Minkowski spacetime (as used in the study of Special Relativity) which is closer in spirit to Hilbert's axiomatic approach to Euclidean geometry than to the vector space approach employed by Minkowski.
We present a brief overview of the axioms as well as of a formalisation of theorems relating to linear order. Proofs and excerpts of Isabelle/Isar scripts are discussed, with a focus on the use of symmetry and reasoning ``without loss of generality''.
\end{abstract}
\section{Problem and Motivation}

The special theory of relativity (SR) \cite{einstein1905} is a canonical part of modern physics. It is customarily taught in the formulation first established by Minkowski in 1908 \cite{minkowski1908}, which is largely determined by the geometry called \emph{Minkowski spacetime}.
The key to understanding SR, of interest both paedagogically and philosophically, is thus to reduce Minkowski spacetime to its essentials, and glean intuition from these building blocks and their use.
This is the aim of three decades of work by John Schutz \cite{schutz1973,schutz1981},
culminating in a categorical system of fifteen independent axioms in 1997 \cite{schutz1997}.
Interestingly, Schutz' axioms depart from the coordinate formalism
and instead imposes
geometrical constraints similar to Hilbert's \emph{Grundlagen} \cite{hilbert1950} onto primitives that can be interpreted as inertial particle trajectories.

Our aim is to formalise Schutz' system of axioms
published in 1997 \cite{schutz1997}
and several of its theorems in the interactive theorem prover Isabelle/HOL.%
\footnote{The formalisation is accessible on the Archive of Formal Proofs: \href{https://www.isa-afp.org/entries/Schutz_Spacetime.html}{https://www.isa-afp.org/entries/Schutz\_Spacetime.html}.}
The beginning of this work has been reported on at ADG in 2018 \cite{palmer2018}. Formalisation of thirteen theorems has now been completed, including many helper lemmas not included in the manuscript, requiring thus far only minor changes to Schutz' axioms.
As one goal of Schutz' monograph is to build a model of his axioms, each verified theorem increases our confidence in the consistency and sufficiency (for Schutz' and our purposes) of the axioms of order and incidence.
Beyond ensuring the correctness of proofs, a mechanisation can provide additional insight into the axioms and their use, and has already begun to do so for us. For example, beyond fundamental correctness issues, the identification of patterns of reasoning that are used repeatedly throughout the formalisation, but not highlighted or mentioned in the original prose, point to intuitively understood facts that are not obvious from a purely logical point of view.

We have noticed that Schutz' prose reasoning becomes more laborious to mechanise as more intuition about the real number line is used. A particular example is the use of reasoning about orderable sets of elements, which are often assumed ``without loss of generality'' to be in some specific configuration. We will emphasise this example when presenting our work (see Sec.~\ref{sec:wlog}).

\section{Background: Axioms for Spacetime}

Several axiom systems have been proposed for Minkowski spacetime. The most widespread use is made of a formulation of spacetime as a real vector space equipped with a bilinear form called the ``Minkowski metric", but more geometric approaches have also been studied.

Schutz proposes several iterations of his own axiom system, starting with a formulation based on primitive particles and the binary \textit{signal} relation in 1973 \cite{schutz1973}. The next iteration in 1981 replaces \textit{signals} with a binary \textit{temporal order} relation, and light signals become an entirely derived notion, whose existence is proven, not assumed \cite{schutz1981}. It is the final axiom system, published in a monograph in 1997 \cite{schutz1997}, that is of interest to us: it contains many of the axioms of earlier systems as theorems, while also boasting the property of independence (see Sec.~\ref{sec:axioms} for details).
Systems formulated by Szekeres \cite{szekeres1968} and Walker \cite{walker1959} also rely on undefined bases and axioms inspired by physical intuition, and Schutz cites them as direct predecessors to his work.
Another early approach is that of Robb \cite{robb1936}, based on events and an ordering relation, and continued by Mundy \cite{mundy1986,mundy1986a}.
A first-order alternative to Schutz is given by Goldblatt \cite{goldblatt1989,goldblatt2012}, who relies on a relation of orthogonality in addition to the betweenness Schutz employs in his system of 1997.

More recently, an extension of Tarski's Euclidean ideas to Goldblatt's approach to Minkowski spacetime was given by Cocco and Babic \cite{cocco2020}. Their system is mostly formulated in first-order logic, but with a second-order continuity axiom in order to show the usual four-dimensional Minkowski spacetime is a model.
A flexible first-order system of axioms describing several different theories of relativity was given by Andr\'eka et al. \cite{andreka2011,andreka2013}. Notably, there exists a mechanisation of this approach in Isabelle/HOL by Stannett and N\'emeti \cite{stannett2014}. In contrast to what we propose here, Stannett and N\'emeti assume an underlying coordinate formulation and use first-order axioms, while Schutz' system is second-order, and his Isomorphism Theorem linking it to the usual coordinate model is one of his final results.

For a comprehensive picture of interactive and automated work in geometry, we refer to a recent review \cite{narboux2018}. Here we just note a few appearances of automated reasoning about symmetry or ``without loss of generality'' (WLOG).
Harrison dedicates an entire paper to WLOG reasoning in HOL Light \cite{harrison2009}.
During a mechanisation of Hilbert's \emph{Grundlagen} \cite[sec.~10.4.3]{scott2015}, Scott argues for a procedural script to repeat proofs for multiple symmetric cases, and thus circumvents the introduction of WLOG lemmas.
Meikle and Fleuriot consider the problem of translation invariance in automated geometry \cite[sec.~5]{meikle2010}. Translation invariance can be seen as the freedom to choose an \emph{origin} freely, while our symmetry (in Sec.~\ref{sec:wlog}) allows free choice of \emph{direction} for chains.


\section{The Axioms of Schutz}\label{sec:axioms}

We reproduce here (with minor cosmetic changes to Definitions~\ref{def:2} and \ref{def:3}), for the sake of completeness, the axioms as Schutz gives them in his monograph \emph{Independent Axioms for Minkowski Spacetime}, including any needed definitions \cite[pp.~9-17]{schutz1997}. Brevity precludes a reproduction of our mechanisation here, and we refer to a prior contribution \cite{palmer2018} and our recent, detailed description \cite{schmoetten2021} for details.

Our undefined primitives are a set $\E$, whose elements are called \emph{events}, and a set $\P$ of sets of events. Elements of $\P$ are called \emph{paths}. The only other primitive we need is a ternary relation defined on $\E$, which is called \emph{betweenness}, and denoted $\ord{\cdot}{\cdot}{\cdot}$ in prose.

Schutz lays out his axioms in two main groups: order and incidence, a separation introduced in Hilbert's \textit{Grundlagen der Geometrie} \cite{hilbert1950}. The first group relates betweenness to events and paths, and establishes a kind of plane geometry with axiom O6. It includes the symmetry of betweenness in Axiom~\ref{ax:O2}, as well as ideas similar to transitivity and irreflexivity.

\begin{axiom}[\customlabel{ax:O1}{O1}]
For events $a,b,c \in \E$,
\[\ord{a}{b}{c} \implies \exists Q \in \P : a,b,c \in Q.\]
\end{axiom}
\begin{axiom}[\customlabel{ax:O2}{O2}]
For events $a,b,c \in \E$,
\[\ord{a}{b}{c} \implies \ord{c}{b}{a}.\]
\end{axiom}
\begin{axiom}[\customlabel{ax:O3}{O3}]
For events $a,b,c \in \E$,
\[\ord{a}{b}{c} \implies a,b,c \text{ are distinct.}\]
\end{axiom}
\begin{axiom}[\customlabel{ax:O4}{O4}]
For distinct events $a,b,c,d \in \E$,
\[\ord{a}{b}{c} \text{ and } \ord{b}{c}{d} \implies \ord{a}{b}{d}\;.\]
\end{axiom}
\begin{axiom}[\customlabel{ax:O5}{O5}]
For any path $Q \in \P$ and any three distinct events $a,b,c \in Q$,
\[
    \ord{a}{b}{c} \;\text{ or }\; \ord{b}{c}{a} \;\text{ or }\; \ord{c}{a}{b} \;\text{ or } \\
    \ord{c}{b}{a} \;\text{ or }\; \ord{a}{c}{b} \;\text{ or }\; \ord{b}{a}{c} \;.
\]
\end{axiom}

\noindent
Schutz now defines \emph{chains}, which appear throughout our formalisation. He claims this is to preserve the independence of his axioms, and specifically, the independence of his version of the axiom of Pasch, here numbered as Axiom~\ref{ax:O6}.

\begin{definition}[Chain]\label{def:1}
A sequence of events $\;Q_0, Q_1, Q_2,\; \dots\;$ (of a path $Q$) is called a \emph{chain} if:
\begin{enumerate}
\item[(i)] it has two distinct events, or
\item[(ii)] it has more than two distinct events and for all $i \geq 2$, \[\ord{Q_{i-2}}{Q_{i-1}}{Q_{i}}\;.\]
\end{enumerate}
\end{definition}

\begin{axiom}[\customlabel{ax:O6}{O6}]
If $Q$, $R$, $S$ are distinct paths which meet at events $a \in Q \cap R$, $b \in Q\cap S$, $c \in R \cap S$ and if:
\begin{enumerate}
    \item[(i)] there is an event $d \in S$ such that $\ord{b}{c}{d}$, and
    \item[(ii)] there is an event $e \in R$ and a path $T$ which passes through both $d$ and $e$ such that $\ord{c}{e}{a}$,
\end{enumerate}
then $T$ meets $Q$ in an event $f$ which belongs to a finite chain $\chain[\cdot\cdot]{a,f,b}$.
\end{axiom}

\begin{figure}
\centering
\begin{subfigure}[b]{0.48\textwidth}
    \centering
\baselineskip=10pt

\begin{tikzpicture}[line cap=round,line join=round,>=triangle 45,x=0.5cm,y=0.3cm]
\clip(3,-9) rectangle (14.5,5);
\draw [line width=1pt,domain=2:14.5] plot(\x,{(--72-8*\x)/10});
\draw [line width=1pt,domain=2:14.5] plot(\x,{(--42-1*\x)/-7});
\draw [line width=1pt,domain=2:14.5] plot(\x,{(-48--9*\x)/-3});
\draw [line width=1pt,color=tikz-darkblue,domain=2:14.5] plot(\x,{(-35.2--3.3*\x)/1.1});
\begin{normalsize}
\draw [fill=black] (4,4) circle (1.5pt);
\draw[color=black] (4.3,4.4) node {$b$};
\draw [fill=black] (14,-4) circle (1.5pt);
\draw[color=black] (14.2,-3.3) node {$a$};
\draw[color=black] (6.8,2.6) node {$Q$};
\draw [fill=black] (7,-5) circle (1.5pt);
\draw[color=black] (7.2,-4.5) node {$c$};
\draw[color=black] (11.5,-5.1) node {$R$};
\draw[color=black] (5.2,-1.7) node {$S$};
\draw [fill=tikz-blue] (8,-8) circle (2.5pt);
\draw[color=tikz-blue] (8.6,-7.8) node {$d$};
\draw [fill=tikz-blue] (9.1,-4.7) circle (2.5pt);
\draw[color=tikz-blue] (9.4,-5.3) node {$e$};
\draw[color=tikz-darkblue] (11.7,4.061061061061062) node {$T$};
\draw [fill=tikz-red] (10.315789473684209,-1.052631578947368) circle (2pt);
\draw[color=tikz-red] (10.75,-0.5) node {$f$};
\end{normalsize}
\end{tikzpicture}
    \caption{\label{fig:O6}Intuitive visualisation of axiom O6. A path $T$ that meets $S$ externally to the triangle $QRS$ (in $d$) and meets $R$ internally (in $e$), must meet the third side of the triangle internally (in $f$).}
\end{subfigure}
\hfill
\begin{subfigure}[b]{0.48\textwidth}
    \centering
\baselineskip=10pt

\begin{tikzpicture}[line cap=round,line join=round,>=triangle 45,x=0.5cm,y=0.3cm]
\clip(2,-9) rectangle (14.5,5);
\draw [line width=1pt,domain=2:14.5] plot(\x,{(--72-8*\x)/10});
\draw [line width=1pt,domain=2:14.5] plot(\x,{(--42-1*\x)/-7});
\draw [dashed,line width=1pt,opacity=0.5] (7,-5) -- (9,0); 
\draw [dashed,line width=1pt,opacity=0.5] (10,-4.6) -- (9,0); 
\begin{normalsize}
\draw[color=black] (6.8,2.6) node {$R$};
\draw[color=black] (3.,-6.5) node {$Q$};
\draw [fill=black] (14,-4) circle (1.5pt);
\draw[color=black] (14.1,-3.2) node {$x$};
\draw [fill=black] (7,-5) circle (1.5pt);
\draw[color=black] (7.2,-6) node {$Q_a$};
\draw [fill=black] (9,0) circle (1.5pt);
\draw[color=black] (9.3,0.8) node {$R_w$};
\draw [fill=tikz-darkblue] (10,-4.6) circle (1.5pt);
\draw[color=tikz-darkblue] (10.2,-5.6) node {$Q_y$};
\end{normalsize}
\end{tikzpicture}
    \caption{\label{fig:unreach_via}The event $Q_y$ belongs to the unreachable subset of $Q$ from $Q_a$ via $R$. Thus there is an event $R_w$, such that there are no paths connecting $(Q_a, R_w)$ or $(Q_y, R_w)$ (dashed lines). In this case, $R_w$ also belongs to the unreachable subset of $R$ from $Q_a$.}
\end{subfigure}
\caption{Two example visualisations in the real plane: axiom~O6 (similar to the axiom of Pasch) and unreachable subsets.}
\end{figure}
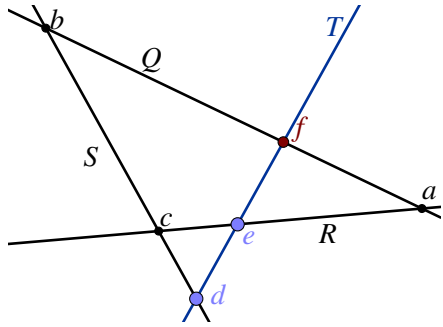
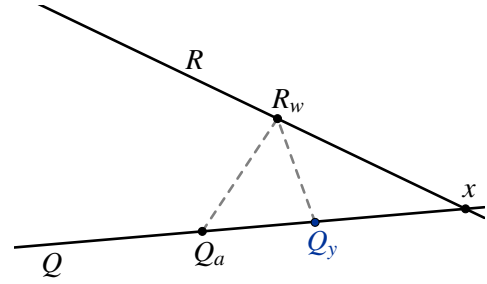

\noindent
Schutz' axioms of order are strongly inspired by the second group of axioms given by Hilbert in his \emph{Grundlagen der Geometrie} \cite{hilbert1950}. Hilbert's Axiom~II.1 combines Schutz' Axioms~\ref{ax:O1}, \ref{ax:O2}, \ref{ax:O3}; Hilbert's II.2 becomes Schutz' Theorem~6, II.3 becomes Theorem~1. Pasch's axiom exists in both systems, respectively as II.4 and (in a slightly different formulation) \ref{ax:O6} (see also Fig.~\ref{fig:O6}).

The second group, the axioms of incidence, deals with the relationships between events and paths, and also contains statements regarding unreachable subsets, which make a model with only Galilean relativity impossible.

\begin{axiom}[\customlabel{ax:I1}{I1}] 
$\E$ is not empty.
\end{axiom}
\begin{axiom}[\customlabel{ax:I2}{I2}] 
For any two distinct events $a,b \in \E$ there are paths $R$, $S$ such that $a \in R$, $b\in S$, and $R\cap S\neq \emptyset$.
\end{axiom}
\begin{axiom}[\customlabel{ax:I3}{I3}] 
For any two distinct events, there is at most one path which contains both of them.
\end{axiom}

\noindent
We omit here the Axiom of Dimension, which fixes the dimension (in a meaning similar to that of linear algebra) of possible models. It relies on a fairly complicated definition of path (in)dependence, and adds little to the discussion in Sec.~\ref{sec:wlog}. Thus we advance to unreachable subsets (see also Fig.~\ref{fig:unreach_via}), which are essential for establishing the universal speed limit that is central to SR.

\begin{definition}[Unreachable Subset from an Event]\label{def:2}
Given a path $Q$ and an event $b \notin Q$, we define the unreachable subset of $Q$ from $b$ to be
\[Q(b,\emptyset) := \left\lbrace x \in Q \text{ such that there is no path which contains $b$ and $x$} \right\rbrace \;.\]
\end{definition}

\begin{definition}[Unreachable Subset via a Path]\label{def:3}
For any two distinct paths $Q$, $R$ which meet at an event $x$, we define the event $Q_y \in Q$ as unreachable from $Q_a$ via $R$ if $\ord{x}{Q_y}{Q_a}$ and there is an event $R_w \in R$ such that $Q_a, Q_y \in Q(R_w,\emptyset)$. That is, the unreachable subset of $Q$ from $Q_a$ via $R$ is
\begin{align*}
    Q(Q_a,R,x,\emptyset) := \lbrace Q_y :\; 
    & \ord{x}{Q_y}{Q_a} \land \\
    & \exists R_w \in R.\; (Q_a \in Q(R_w,\emptyset) \land Q_y \in Q(R_w,\emptyset)) \rbrace .
\end{align*}
\end{definition}

\noindent
Axioms \ref{ax:I6}--\ref{ax:I7} are used later to show that unreachable sets are connected and bounded. An event \hbox{$a\in Q(b,\emptyset)$} is called \emph{unreachable} (from $b$); all other events on $Q$ are termed \emph{reachable}.

\begin{axiom}[\customlabel{ax:I5}{I5}]
For any path $Q$ and any event $b \notin Q$, the unreachable set $Q(b,\emptyset)$ contains (at least) two events.
\end{axiom}

\begin{axiom}[\customlabel{ax:I6}{I6}]
Given any path $Q$, any event $b \notin Q$ and distinct events $Q_x, Q_z \in Q(b,\emptyset)$, there is a finite chain $[Q_0 \;\dots\; Q_n]$ with $Q_0 = Q_x$ and $Q_n = Q_z$ such that for all $i \in \left\lbrace1,2,\dots,n\right\rbrace$,
\begin{enumerate}
    \item[(i)] $Q_i \in Q(b,\emptyset)$
    \item[(ii)] $\ord{Q_{i-1}}{Q_y}{Q_i} \implies Q_y \in Q(b,\emptyset)$.
\end{enumerate}
\end{axiom}

\begin{axiom}[\customlabel{ax:I7}{I7}]
Given any path $Q$, any event $b \notin Q$, and events $Q_x \in Q \setminus Q(b,\emptyset)$ and $Q_y \in Q(b,\emptyset)$, there is a finite chain
\[\chain{Q_0,Q_m,Q_n}\] 
with $Q_0 = Q_x$, $Q_m = Q_y$ and $Q_n \in Q \setminus Q(b,\emptyset)$.
\end{axiom}

\noindent
Two further axioms, of Isotropy and Continuity, are omitted here for brevity. They have no bearing on the content of this abstract. More detail can be found, in addition to the original monograph \cite{schutz1997}, in an earlier contribution to ADG \cite{palmer2018}.



Given the correspondences of axioms noted above, our formalisation bears some similitude to work by Meikle, Scott and Fleuriot on formalising Hilbert's \emph{Grundlagen der Geometrie} in Isabelle \cite{meikle2003,scott2008} and HOL Light \cite{scott2011,scott2015}. For example, our definition of chains (Def.~\ref{def:1}), one of the most fundamental constructs in this mechanisation, relies on an adapted definition from Scott's work on the \emph{Grundlagen}.
As another example, we employ the same weakening of Schutz' Axiom~O3 that can be found in Scott's formalisation of Hilbert's Axiom~II.1. Scott also formalises a proof of uniqueness (up to reversal) of chains (this appears briefly in Section~\ref{sec:wlog}): while he obtains it from a remark of Hilbert's \cite[Section~6.7.2]{scott2015}, we derived it by necessity in an early version of our proof of Theorem~13, and found the correspondence only later.
Notice the formalisations of Hilbert's \emph{Grundlagen} cited here focus on the first three groups of axioms, which exclude the parallel and continuity axioms.

\section{Theorems and WLOG}\label{sec:thms}

The complete formalisation is over nine thousand lines long. Schutz' admirably detailed account (for prose) covers 22 pages. Estimating thirty lines on each page, this leaves us with a de Bruijn factor \cite{wiedijk2000,debruijn1994a} of roughly 14. This is not exceptional: while many formalisations only report de Bruijn factors as low as 3 to 6, values above 20 can be found \cite{dzamonja2020}. One should note that the axiomatisation by itself would have a factor of only around 4. The thirteen formalised theorems and their proofs, together with most added intermediate lemmas, have de Bruijn factor of roughly 23. This, we estimate, is largely due to the later proofs of the chapter relying more strongly on Schutz' geometric intuition. Thus our formal constructions had to become more and more elaborate, the prime example being our collection of WLOG lemmas (Sec.~\ref{sec:wlog}).


Fourteen theorems are proven in Schutz' chapter on \emph{Temporal order on a path}. We have completed the formalisation of thirteen of them, but give here only a short summary of the results and their proofs. We will then go into more detail concerning crucial steps in the proofs of three results that rely on lemmas of symmetry. This section will feature excerpts from our Isabelle/HOL mechanisation. Betweenness is denoted using double brackets \lstinline|[[_ _ _]]| in Isabelle; this choice is made because single brackets (used in Schutz' prose) represent lists.

Chains play an important role here, as they are the main way of extending betweenness to more than three events. Schutz' formulation (given before Axiom~\ref{ax:O6}) needs to be made more explicit in the mechanisation. In fact, the function $\mathbb{N} \rightarrow \E$ that is implicit in Schutz' indices needs to be kept track of. It is this function $f$ that must satisfy the conditions given by Schutz, such as the line below (to be contrasted with Def.~\ref{def:1}).
\enlargethispage{2\baselineskip}
\begin{lstlisting}
    \<forall>n n' n''. n<n' \<and> n'<n'' \<longrightarrow> [[(f n) (f n') (f n'')]]"
\end{lstlisting}
\noindent
Together with several conditions that enable working with indices, this function $f$ then allows us to define a chain \lstinline|[f[a..b]Q]| in Isabelle, where $Q$ is the set of all events of the chain, and the events $a,b$ have ``indices'' $0$, $|Q|-1$ respectively (so that $f(0)=a=Q_0$ in Schutz' notation).

\subsection{Summary of the Mechanised Results}

Theorem 1 establishes the symmetry of Axiom~\ref{ax:O2} as the only symmetry (i.e. $\ord{a}{b}{c}$ implies $\ord{c}{b}{a}$ and falsifies all other orderings of $a,b,c$). Theorem 2 proves that the indexing function of a chain is an order-preserving injection. Proof of the first was straightforward though it required adjusting Axiom~\ref{ax:O4} (see Palmer and Fleuriot at ADG 2018 \cite{palmer2018}), while the second, a proof by induction, is more involved, but follows the proof given by Schutz precisely.

\begin{figure}
    \centering
    \begin{subfigure}[t]{0.48\textwidth}
        \centering


\definecolor{yqqqqq}{rgb}{0.5019607843137255,0,0}
\definecolor{qqttzz}{rgb}{0,0.2,0.6}
\definecolor{xdxdff}{rgb}{0.49019607843137253,0.49019607843137253,1}

\begin{tikzpicture}[line cap=round,line join=round,>=triangle 45,x=0.5cm,y=0.3cm]
\clip(2,-9) rectangle (14.5,5);
\draw [line width=1pt,domain=2:14.5] plot(\x,{(--72-8*\x)/10});
\draw [line width=1pt,domain=2:14.5] plot(\x,{(--42-1*\x)/-7});
\draw [line width=1pt,domain=2:14.5] plot(\x,{(-48--9*\x)/-3});
\draw [line width=1pt,color=qqttzz,domain=2:14.5] plot(\x,{(-35.2--3.3*\x)/1.1});
\begin{normalsize}
\draw [fill=black] (4,4) circle (1.5pt);
\draw[color=black] (4.3,4.4) node {$b$};
\draw [fill=black] (14,-4) circle (1.5pt);
\draw[color=black] (14.2,-3.3) node {$a$};
\draw [fill=black] (7,-5) circle (1.5pt);
\draw[color=black] (7.2,-4.5) node {$c$};
\draw [fill=xdxdff] (8,-8) circle (2.5pt);
\draw[color=xdxdff] (8.6,-7.8) node {$d$};
\draw [fill=xdxdff] (9.1,-4.7) circle (2.5pt);
\draw[color=xdxdff] (9.4,-5.3) node {$e$};
\draw [fill=yqqqqq] (10.315789473684209,-1.052631578947368) circle (2pt);
\draw[color=yqqqqq] (10.75,-0.5) node {$f$};
\end{normalsize}
\end{tikzpicture}

        \caption{The First and Second Collinearity Theorems. Provided a kinematic triangle $abc$ and a path $de$ with
        $[aec]$ and $[bcd]$, we obtain an event $f$ with $[afb]$ and $[def]$.
        }
        \label{fig:collinearity}
    \end{subfigure}
    \hfill
    \begin{subfigure}[t]{0.48\textwidth}
        \centering
\baselineskip=10pt

\definecolor{yqqqqq}{rgb}{0.5019607843137255,0,0}
\definecolor{qqttzz}{rgb}{0,0.2,0.6}
\definecolor{xdxdff}{rgb}{0.49019607843137253,0.49019607843137253,1}

\begin{tikzpicture}[line cap=round,line join=round,>=triangle 45,x=0.5cm,y=0.3cm]
\clip(-2,-2) rectangle (11,11);
\draw[line width=1pt] (-2,-2) -- (0,0) -- (7,7) -- (10,10);
\draw[dashed,line width=1pt,opacity=0.5] (1,8) -- (5,5);
\draw[dashed,line width=1pt,opacity=0.5] (1,8) -- (0.5,0.5);
\draw[dashed,line width=1pt,opacity=0.5] (6,1) -- (2,2);
\draw[dashed,line width=1pt,opacity=0.5] (6,1) -- (6.5,6.5);
\begin{normalsize}
\draw[color=black] (10,9) node {$Q$};
\draw[color=black,opacity=0.5] (5,3) node {$Q(a,\emptyset$)};
\draw[color=black,opacity=0.5] (2.5,4.5) node {$Q(b,\emptyset$)};
\draw [fill=black] (1,8) circle (1.5pt);
\draw[color=black] (0.8,8.6) node {$b$};
\draw [fill=black] (6,1) circle (1.5pt);
\draw[color=black] (6.4,0.8) node {$a$};
\draw [fill=qqttzz] (0,0) circle (1.5pt);
\draw[color=qqttzz] (0.4,-0.8) node {$y$};
\draw [fill=qqttzz] (7,7) circle (1.5pt);
\draw[color=qqttzz] (7.2,6.2) node {$z$};
\end{normalsize}
\end{tikzpicture}
        \caption{Theorem \ref{thm:14}(i). Both events $a$ and $b$ must be reachable from the path $Q$ in order to obtain bounding events $y,z$.}
        \label{fig:thm14i}
    \end{subfigure}
    \caption{Visualisation of the Theorems of Collinearity and Existence.}
    \label{fig:col-ex}
\end{figure}
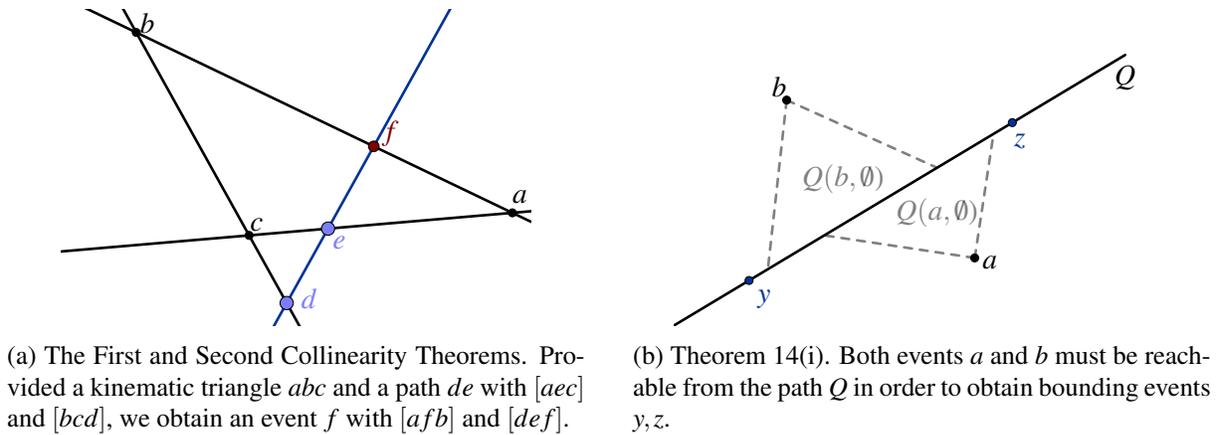

The First and Second Collinearity Theorems (numbered 3 and 7) make use of \emph{kinematic triangles} (triplets of distinct events such that any pair lies on one of three distinct paths) to establish elements of a plane-like geometry on the set of events. Thus they are most easily visualised pictorially (Fig.~\ref{fig:collinearity}). 
Both proofs follow Schutz' prose easily, with only a few additional steps needed to establish ``obvious'' properties of kinematic triangles.

Given a reachable event $a$ and an unreachable event $b$ on a path $ab$, Theorem~4 obtains a reachable event $c$ with $\ord{a}{b}{c}$, i.e. it ``bounds'' a given unreachable event.
Since it follows quite clearly from Theorem 2 and Axiom \ref{ax:I7}, this theorem is verified straightforwardly in Isabelle. Once we add Theorem 13, which proves that unreachable sets are connected, unreachable sets become somewhat reminiscent of intervals on the real number line. This theorem's proof is more involved, relying on Axiom \ref{ax:I6} and Theorem 11. In fact, we had to amend our formalisation of Axiom \ref{ax:I6} to include explicit treatment of two-event chains, where the indexing function makes no implication about the ternary betweenness relation. It also necessitated proof of additional results relating to symmetry and uniqueness (see Sec.~\ref{sec:wlog}), whose statement or use are not in Schutz' manuscript.

Theorem 6, the Prolongation Theorem, obtains an event $c\in Q$ with $\ord{a}{b}{c}$ for any distinct events $a,b$ on any path $Q$. Thus it also implies that paths have at least the cardinality of the natural numbers, i.e. paths contain infinitely many events. Notice the Prolongation Theorem makes no statement about density. Since its formalisation is readable and succinct, we give it below.

\begin{lstlisting}
theorem (*6i*) prolong_betw:
  assumes path_Q: "Q \<in> \<P>" and "a \<in> Q" "b \<in> Q" "a \<noteq> b"
  shows "\<exists>c\<in>\<E>. [[a b c]]"
  
theorem (*6ii*) infinite_paths:
  assumes "P\<in>\<P>"
  shows "infinite P"
\end{lstlisting}

Theorems 8 and 9, as well as the three lemmas preceding the proof of Theorem 9, are stepping stones towards Theorem 10, part of whose proof is examined below. Theorem 8 is an opposite analogue to the Collinearity Theorems in a way: it shows that no path may cross all three sides of a triangle internally; it is a statement about non-collinearity. It serves to prove by contradiction that the betweenness relation satisfies a kind of transitivity, expressed in lemmas such as the one below. We add several similar results to Schutz' three. While these additional lemmas are, strictly speaking, redundant after Theorem 10, they make proofs easier in Isabelle, as they can avoid the explicit creation of chains when reasoning about betweenness.
\begin{lstlisting}
lemma abc_acd_bcd:
  assumes abc: "[[a b c]]" and acd: "[[a c d]]"
  shows "[[b c d]]"
\end{lstlisting}

\noindent
The proof of the first such Lemma follows Schutz in general, although the mechanisation treats the consideration of different cases in a slightly different way to the prose proof. Using this, the remaining lemmas are straightforward to prove, as is Theorem 9.
Mechanising Schutz' proof for Theorem 8 is straightforward, with the caveat that three different orderings of events must be considered. Schutz considers one of them, and argues the remaining ones follow by ``cyclic interchange of the symbols [labelling the events] throughout the proof''  \cite[p.~23]{schutz1997}. In Isabelle, this was done explicitly, by copying the proof and replacing the variable names. It is the first appearance of \emph{without loss of generality} (WLOG) reasoning, which we will expand on in Sec.~\ref{sec:wlog}.

Theorem 10 (see also Sec.~\ref{sec:wlog}) generalises Theorem 9 and its helper lemmas by stating that any finite set of events on a path forms a chain. The proof is by induction on the cardinality of the set of events. The induction step is split into three cases: given an event-set $X$, which is a chain by the inductive hypothesis, we can add an event on either side (index $0$ or $|X|$), or in the interior of the chain. The two side-cases are clearly symmetric, and will be treated below. We only comment here that several additional steps were needed in Isabelle as compared to the prose, to prove existential statements that are fairly trivial to the inspecting mathematician.

Theorem 11 allows us to split up a path into a set of segments and two rays. The terminology here is that of the real number line; Schutz' definitions are a little more technical, but the intended correspondence is clear. This clarity is a two-edged sword: it makes it easy to talk about results in an intuitive way, but also invites hasty, faulty statements. Thus we had to weaken Schutz' statement of Theorem 11, as he made unproveable claims about the number of segments obtained. His claim becomes proveable only during the next chapter of his monograph, and follows from the as of yet unjustified identification of paths with real number lines.
Similar issues lead to a much larger ratio between lengths of prose and Isabelle proofs for the later theorems we mechanised, as compared to the earlier ones.

Theorems 5 and 14 (see Fig.~\ref{fig:thm14i}), finally, form a unit much like Theorems 3 and 7. The First and Second Existence Theorems provide the starting point of several geometric follow-on proofs by obtaining events and paths in specific configurations. The First is mechanised easily by following Schutz, while the Second, for the reason outlined in the paragraph above, takes some extra work. It is here that we make use of explicit WLOG-theorems to obtain sufficient subgoals and avoid explicitly considering all possible orderings of four events.

\subsection{Symmetry and WLOG}\label{sec:wlog}

The second sentence in Schutz' statement of Theorem 10 highlights the central problem of this section.

\setcounter{theorem}{9}
\begin{theorem}\label{thm:10} 
Any finite set of distinct events of a path forms a chain. That is, any set of $n$ distinct events can be represented by the notation $a_1,a_2,\dots,a_n$ such that\vspace{-7pt}
\[[a_1 \; a_2 \dots a_n]\;.\]
\end{theorem}

\noindent
Arbitrary choice of a naming scheme is often hard to emulate in Isabelle proof script. Instead, our mechanised Theorem 10 only talks about a property of a set of events. In the listing below, \lstinline|ch X| is an abbreviation that means ``either $X$ is a two-event subset of a path or there exists a function $\mathbb{N}\rightarrow\E$ that orders $X$ into a chain'' (cf. Def~\ref{def:1}).

\begin{lstlisting}
theorem path_finsubset_chain:
  assumes "Q \<in> \<P>" and "X \<subseteq> Q" and "card X \<ge> 2"
  shows "ch X"
\end{lstlisting}

\noindent
It is relatively easy to just ignore the second sentence of Schutz's enunciation of Theorem 10, as it adds nothing to the conclusion, but similar phrasing in his reasoning is hard to replicate, and makes for lengthy, repetitive proofs. Two types of reasoning device will be investigated here: the reversal symmetry originally encoded in Axiom~\ref{ax:O2}, and the WLOG reasoning linked additionally to arbitrary naming of events. As seen from Theorem 10, this is most noticeable when a property of a structure, such as a set of events, is expressed in terms of elements of that structure, such as the ordering of the set's events. We will see this again when we talk about intervals in terms of their endpoints (Sec~\ref{sec:wlog}).

For now, let us dive into the proof of Theorem 10. As hinted in Sec.~\ref{sec:thms}, the crucial part is to append one new event $b$ on a path $Q$ to a chain $X$ on the same path, and show the resulting set $X\cup\{b\}$ is also a chain. This is done by constructing a suitable indexing function $f$, and showing that $f$ satisfies the betweenness conditions that order $X\cup\{b\}$ into a chain. The event $b$ can be on the interior of $X$, or on either side. The proof for appending an event $b$ on the left edge of a chain $X$ (ordered by $f$) is based on the indexing function $g$ where $g(0) = x$ and $g(i)=f(i-1)$ for valid indices $i>0$, and leads to the following lemma.

\begin{lstlisting}
lemma chain_append_at_left_edge:
  assumes long_ch_X: "[f[a\<^sub>1..a..a\<^sub>n]X]"
      and bX: "[[b a\<^sub>1 a\<^sub>n]]"
    fixes g
  defines g_def: "g \<equiv> (\<lambda>j. if j\<ge>1 then f (j-1) else b)"
    shows "[g[b .. a\<^sub>1 .. a\<^sub>n](insert b X)]"
\end{lstlisting}

\noindent
The proof of this lemma explicitly verifies all the needed betweenness relations (for any triplet of indices), and uses Theorem 2. As this is lengthy, instead of reproducing the proof for the case of a new event at the right edge, we instead use a symmetry result in conjunction with the lemma above.

\begin{lstlisting}
lemma chain_sym:
  assumes "[f[a..b..c]X]"
    shows "[\<lambda>n. f(card X - 1 - n)[c..b..a]X]"
\end{lstlisting}

\noindent
The lemma \lstinline|chain_sym| can be viewed as a generalisation of Axiom~\ref{ax:O2} to chains, but is absent from the monograph. We further prove that the reversed chain is the \emph{only} other consistent way of indexing a chain of events, which can be compared to Axiom~\ref{ax:O5}.
We posit it is the absence of these results that leads to Schutz' very simple proof of Theorem~13, as the two chains he obtains from Theorem~10 and Axiom~\ref{ax:I6} respectively are assumed equal (without justification). Our first attempt at mechanised proof had to consider the reversed case also, but using the above results, we can guarantee that no other cases are possible. %
\footnote{We have since simplified the proof of Theorem~13 by using a version of Theorem~11 that is slightly different to Schutz' formulation; but the result on chain symmetry and uniqueness remains interesting.}


In the case of a new event added at the right edge, the candidate indexing function of the new chain will be $g:\mathbb{N}\rightarrow\E$, with $g(|X|)=b$ and $g(i)=f(i)$ for valid indices $i<|X|$.
We prove the fact \hbox{\lstinline|[g[a\<^sub>1..a\<^sub>n..b](insert b X)]|} given $\ord{a_1}{a_n}{b}$. This is done by obtaining functions $f_2$ and $g_2$, which can be seen as a reversed version of $f$ and $g$: if $f$ indexes a chain ``left-to-right'', $f_2$ counts ``right-to-left''.

\enlargethispage{\baselineskip}
\begin{lstlisting}
  obtain f2 where f2_def: "[f2[a\<^sub>n..a..a\<^sub>1]X]"
      "f2=(\<lambda>n. f (card X - 1 - n))"
    using chain_sym long_ch_X by blast
  obtain g2 where g2_def:
      "g2 = (\<lambda>j::nat. if j\<ge>1 then f2 (j-1) else b)"
    by simp
\end{lstlisting}

\noindent
We show $g_2$ orders $X\cup\{b\}$ into a chain using \lstinline|chain_append_at_left_edge|, and then reverse it again using \lstinline|chain_sym| to get $g_1$, which also orders $X\cup\{b\}$. Finally, we show $g_1=g$. We give below the case $n \leq |X|-1$; the other case is trivial since then $g_1(n) = b = g(n)$. 
$$
g_1(n) =
g_2(|X\cup\{b\}|-1-n) =
f_2(|X|-1-n) =
f(n) =
g(n)$$

\noindent
Thus our proof is still far more explicit than Schutz' (who just says ``The proof for this case is similar to that of Case (i).''), but at least it is not just a reproduction of the same many betweenness proofs. While the lemma \lstinline|chain_sym| is rather simple to state, the machinery involved in the proof is nonetheless considerable: a chain of four equalities of indexing functions.
The WLOG theorems we present in the following are more verbose to state, as they make explicit all the steps needed for a proof.

Isabelle already includes some WLOG results, notably those in the theory of linear orderings. We reproduce one below (where the addition of Isar-style script is ours).


\begin{lstlisting}
lemma linorder_less_wlog:
  assumes "\<And>a b. a < b \<Longrightarrow> P a b"
      and "\<And>a. P a a"
      and "\<And>a b. P b a \<Longrightarrow> P a b"
    shows "P a b"
\end{lstlisting}

\noindent
This already involves multiple levels of reasoning: the idea that ``element labels are arbitrary'' is captured by the meta-quantification over elements \hbox{\lstinline|\<And>a b|} in each individual assumption. We follow the same style in our WLOG lemmas, which capture the symmetry of Axiom~\ref{ax:O2} and \lstinline|chain_sym| while taking care of the many possible permutations of assigning labels to events. As in \lstinline|linorder_less_wlog|, we will be dealing with a symmetric, reflexive predicate, and our lemma will identify those orderings of elements that are sufficient to prove the desired property for all arguments. Since we are interested in intervals on paths, the predicate is a bit more complicated than $P$ above, in that it takes an event-set and two events, and indicates that these two events define an interval equal to the set.\footnote{%
We give here a WLOG lemma specific to intervals. A lemma using generic relations between events and sets of events has been mechanised, from which the one presented here is derived, but we judged it too verbose for inclusion here.%
}
The interval between $a$ and $b$ is defined as $|ab|=\{x : \ord{a}{x}{b}\} \cup \{a,b\}$.



\pagebreak

{\lstset{escapeinside=||} 
\begin{lstlisting}
lemma wlog_interval_endpoints_distinct:
  assumes "\<And>I J. \<lbrakk>is_int I; is_int J; P I J\<rbrakk> \<Longrightarrow> P J I"
          "\<And>I J a b c d. \<lbrakk>I = interval a b; J = interval c d\<rbrakk>
          \<Longrightarrow> |\tikzmark{starta}|(betw4 a b c d \<longrightarrow> P I J) \<and>
              (betw4 a c b d \<longrightarrow> P I J) \<and>
              (betw4 a c d b \<longrightarrow> P I J)"|\tikzmark{enda}|
  shows "\<And>I J Q a b c d.
           \<lbrakk>I = interval a b; J = interval c d;
            I\<subseteq>Q; J\<subseteq>Q; Q\<in>\<P>;
            a\<noteq>b \<and> a\<noteq>c \<and> a\<noteq>d \<and> b\<noteq>c \<and> b\<noteq>d \<and> c\<noteq>d\<rbrakk>
        \<Longrightarrow> P I J"
\end{lstlisting}}
\begin{tikzpicture}[remember picture,overlay]
\draw[red,rounded corners]
  ([shift={(-3pt,2ex)}]pic cs:starta) 
    rectangle 
  ([shift={(11pt,-0.65ex)}]pic cs:enda);
\end{tikzpicture}


\noindent
The first assumption states that the predicate $P$ is symmetric in its two arguments (both of which are intervals), and can be compared to the third assumption of \lstinline|linorder_less_wlog|.
The important part is the triple conjunction, highlighted in red. After introducing some shared context, this conjunction gives the three essentially different cases of orderings of the endpoints of intervals: disjoint, overlapping, and nested intervals. These give sufficient subgoals for proving a property of two intervals by considering their endpoints explicitly. There is a similar lemma for the case where all four endpoints are not distinct.

We use \lstinline|wlog_interval_endpoints_distinct| when proving that the intersection of two intervals is an interval.
We give below a skeleton of the formal proof in the case where all endpoints are distinct.

\begin{lstlisting}
  let ?prop = "\<lambda> I J. is_int (I\<inter>J) \<or> (I\<inter>J) = {}"
  { fix I J a b c d
    assume "I = interval a b" "J = interval c d"
    { assume "betw4 a b c d"
      have "I\<inter>J = {}" (*...*)
    } { assume "betw4 a c b d"
      have "I\<inter>J = interval c b" (*...*)
    } { assume "betw4 a c d b"
      have "I\<inter>J = interval c d" (*...*)
    } }
\end{lstlisting}

\noindent
The overall proof structure is clear: all the needed cases are considered, and shown to lead to \lstinline|?prop I J|. Proving the final statement is then simple, using the WLOG lemma we have given above.

\begin{lstlisting}
  then show "is_int (I1\<inter>I2)"
    using wlog_interval_endpoints_distinct symmetry assms
    by simp
\end{lstlisting}

{
}

\noindent
The proof of Theorem 14 uses more generic WLOG lemmas, where the \lstinline|interval| function is replaced with any function of one set (of events) and two events, all of which lie on a path, and symmetric in the two event-arguments. The prose for the first part of Theorem 14 is reproduced here; see also Fig.~\ref{fig:thm14i}.

\setcounter{theorem}{13}
\begin{theorem}[Second Existence Theorem, part (i)]\label{thm:14}
Given a path $Q$ and a pair of events $a,b \notin Q$, each of which can be joined to $Q$ by some path, there are events $y,z \in Q$ such that
\[\ord{y}{Q(a,\emptyset)}{z} \text{ and } \ord{y}{Q(b,\emptyset)}{z}\;.\]
\end{theorem}

\noindent
In the final line, betweenness is extended to sets, so that for a set $S$,
\[\ord{a}{S}{b} \iff \forall x \in S: \ord{a}{x}{b}\;.\]
The proposition to be proven is then ``the union of the unreachable sets $Q(a,\emptyset)$ and $Q(b,\emptyset)$ is bounded'', which can be compared to \lstinline|?prop| in the proof above, or to \lstinline|P| in \lstinline|wlog_interval_endpoints_distinct|. Given that the proof proceeds by first finding bounds on either set (similarly to having endpoints for each of two intervals), and then proves the statement about a union (which, much like the interval intersection, is symmetric), the parallels with the proof for intersections of intervals above are clear. Consequently, the proof proceeds in similar manner, with the WLOG lemma used to split up an overall goal into three subgoals (in the case where all bounds are distinct; a further seven are required for various cases of degeneracy), a much more succinct number than the twelve cases that would be obtained by considering the symmetry of \lstinline|chain_sym| alone.

\section{Conclusion}
We have given a brief overview of a formalisation of the axioms and thirteen theorems of Schutz' \emph{Independent Axioms for Minkowski Spacetime} \cite{schutz1997}. With the exception of the Theorem of Continuity, this concludes the third chapter of his monograph, which deals with \emph{Temporal order on a path}. This gives us confidence in the current formulation of the axioms of order and incidence. We hope to extend the mechanisation to include the Axiom and Theorem of Continuity in the near future, reinforcing the analogy between paths and real number lines. Work is also planned on the fourth chapter, which introduces \emph{optical lines} and \emph{causality}.

In addition to a summary of our work, we have given a brief overview of the importance of reasoning about symmetry and arguments ``without loss of generality''. Considering the ease with which these devices are used in prose, and the frequency of their appearance throughout mathematics, we believe future work on automating or simplifying these procedures would be of great benefit. This is a challenging task: multiple layers of reasoning interact, leading to verbose, unwieldy lemmas, a problem which worsens as the lemma in question becomes more generic. The problem is interesting also because WLOG-style reasoning often relies on the mathematician's intuition about a symmetry rather than fully formulated arguments, making it naturally troublesome in mechanisation.



%
%
%
\bibliographystyle{eptcs}
\bibliography{mybibliography}

\begin{thebibliography}{10}
\providecommand{\bibitemdeclare}[2]{}
\providecommand{\surnamestart}{}
\providecommand{\surnameend}{}
\providecommand{\urlprefix}{Available at }
\providecommand{\url}[1]{\texttt{#1}}
\providecommand{\href}[2]{\texttt{#2}}
\providecommand{\urlalt}[2]{\href{#1}{#2}}
\providecommand{\doi}[1]{doi:\urlalt{http://dx.doi.org/#1}{#1}}
\providecommand{\eprint}[1]{arXiv:\urlalt{https://arxiv.org/abs/#1}{#1}}
\providecommand{\bibinfo}[2]{#2}

\bibitemdeclare{article}{andreka2013}
\bibitem{andreka2013}
\bibinfo{author}{Hajnal \surnamestart Andr{\'e}ka\surnameend},
  \bibinfo{author}{Judit~X. \surnamestart Madar{\'a}sz\surnameend},
  \bibinfo{author}{Istv{\'a}n \surnamestart N{\'e}meti\surnameend} \&
  \bibinfo{author}{Gergely \surnamestart Sz{\'e}kely\surnameend}
  (\bibinfo{year}{2013}): \emph{\bibinfo{title}{An {{Axiom System}} for
  {{General Relativity Complete}} with Respect to {{Lorentzian Manifolds}}}}.
\newblock {\sl \bibinfo{journal}{arXiv:1310.1475 [gr-qc]}}.
\newblock \eprint{1310.1475}.

\bibitemdeclare{article}{andreka2011}
\bibitem{andreka2011}
\bibinfo{author}{Hajnal \surnamestart Andr{\'e}ka\surnameend},
  \bibinfo{author}{Istv{\'a}n \surnamestart N{\'e}meti\surnameend},
  \bibinfo{author}{Judit~X. \surnamestart Madar{\'a}sz\surnameend} \&
  \bibinfo{author}{Gergely \surnamestart Sz{\'e}kely\surnameend}
  (\bibinfo{year}{2011}): \emph{\bibinfo{title}{On {{Logical Analysis}} of
  {{Relativity Theories}}}}.
\newblock \eprint{1105.0885}.

\bibitemdeclare{inproceedings}{cocco2020}
\bibitem{cocco2020}
\bibinfo{author}{Lorenzo \surnamestart Cocco\surnameend} \&
  \bibinfo{author}{Joshua \surnamestart Babic\surnameend}
  (\bibinfo{year}{2020}): \emph{\bibinfo{title}{A System of Axioms for
  {{Minkowski}} Spacetime}}.
\newblock In: {\sl \bibinfo{booktitle}{Journal of {{Philosophical Logic}}}}.

\bibitemdeclare{incollection}{debruijn1994a}
\bibitem{debruijn1994a}
\bibinfo{author}{N.~G. \surnamestart {de Bruijn}\surnameend}
  (\bibinfo{year}{1994}): \emph{\bibinfo{title}{A {{Survey}} of the {{Project
  Automath}}}}.
\newblock In \bibinfo{editor}{R.~P. \surnamestart Nederpelt\surnameend},
  \bibinfo{editor}{J.~H. \surnamestart Geuvers\surnameend} \&
  \bibinfo{editor}{R.~C. \surnamestart {de Vrijer}\surnameend}, editors: {\sl
  \bibinfo{booktitle}{Studies in {{Logic}} and the {{Foundations}} of
  {{Mathematics}}}}, {\sl \bibinfo{series}{Selected {{Papers}} on
  {{Automath}}}} \bibinfo{volume}{133}, \bibinfo{publisher}{{Elsevier}}, pp.
  \bibinfo{pages}{141--161}, \doi{10.1016/S0049-237X(08)70203-9}.
\newblock \bibinfo{note}{Reprinted from: Seldin, J. P. and Hindley, J. R.,
  eds., To H. B. Curry: Essays on Combinatory Logic, Lambda Calculus and
  Formalism, pp. 579-606.}

\bibitemdeclare{article}{dzamonja2020}
\bibitem{dzamonja2020}
\bibinfo{author}{Mirna \surnamestart D{\v z}amonja\surnameend},
  \bibinfo{author}{Angeliki \surnamestart {Koutsoukou-Argyraki}\surnameend} \&
  \bibinfo{author}{Lawrence~C. \surnamestart Paulson\surnameend}
  (\bibinfo{year}{2020}): \emph{\bibinfo{title}{Formalising {{Ordinal Partition
  Relations Using Isabelle}}/{{HOL}}}}.
\newblock \eprint{2011.13218}.

\bibitemdeclare{article}{einstein1905}
\bibitem{einstein1905}
\bibinfo{author}{A.~\surnamestart Einstein\surnameend} (\bibinfo{year}{1905}):
  \emph{\bibinfo{title}{Zur {{Elektrodynamik}} Bewegter {{K\"orper}}}}.
\newblock {\sl \bibinfo{journal}{Annalen der Physik}}
  \bibinfo{volume}{322}(\bibinfo{number}{10}), pp. \bibinfo{pages}{891--921},
  \doi{10.1002/andp.19053221004}.

\bibitemdeclare{incollection}{goldblatt1989}
\bibitem{goldblatt1989}
\bibinfo{author}{Robert \surnamestart Goldblatt\surnameend}
  (\bibinfo{year}{1989}): \emph{\bibinfo{title}{First-{{Order Spacetime
  Geometry}}}}.
\newblock In \bibinfo{editor}{Jens~Erik \surnamestart Fenstad\surnameend},
  \bibinfo{editor}{Ivan~T. \surnamestart Frolov\surnameend} \&
  \bibinfo{editor}{Risto \surnamestart Hilpinen\surnameend}, editors: {\sl
  \bibinfo{booktitle}{Studies in {{Logic}} and the {{Foundations}} of
  {{Mathematics}}}}, {\sl \bibinfo{series}{Logic, {{Methodology}} and
  {{Philosophy}} of {{Science VIII}}}} \bibinfo{volume}{126},
  \bibinfo{publisher}{{Elsevier}}, pp. \bibinfo{pages}{303--316},
  \doi{10.1016/S0049-237X(08)70051-X}.

\bibitemdeclare{book}{goldblatt2012}
\bibitem{goldblatt2012}
\bibinfo{author}{Robert \surnamestart Goldblatt\surnameend}
  (\bibinfo{year}{2012}): \emph{\bibinfo{title}{Orthogonality and {{Spacetime
  Geometry}}}}.
\newblock \bibinfo{publisher}{{Springer Science \& Business Media}}.

\bibitemdeclare{incollection}{harrison2009}
\bibitem{harrison2009}
\bibinfo{author}{John \surnamestart Harrison\surnameend}
  (\bibinfo{year}{2009}): \emph{\bibinfo{title}{Without {{Loss}} of
  {{Generality}}}}.
\newblock In \bibinfo{editor}{Stefan \surnamestart Berghofer\surnameend},
  \bibinfo{editor}{Tobias \surnamestart Nipkow\surnameend},
  \bibinfo{editor}{Christian \surnamestart Urban\surnameend} \&
  \bibinfo{editor}{Makarius \surnamestart Wenzel\surnameend}, editors: {\sl
  \bibinfo{booktitle}{Theorem {{Proving}} in {{Higher Order Logics}}}},
  \bibinfo{volume}{5674}, \bibinfo{publisher}{{Springer, Berlin, Heidelberg}},
  pp. \bibinfo{pages}{43--59}, \doi{10.1007/978-3-642-03359-9\_3}.

\bibitemdeclare{book}{hilbert1950}
\bibitem{hilbert1950}
\bibinfo{author}{David \surnamestart Hilbert\surnameend}
  (\bibinfo{year}{1950}): \emph{\bibinfo{title}{The {{Foundations}} of
  {{Geometry}}}}.
\newblock \bibinfo{publisher}{{The Open Court Publishing Company}}.

\bibitemdeclare{inproceedings}{meikle2010}
\bibitem{meikle2010}
\bibinfo{author}{Laura \surnamestart Meikle\surnameend} \&
  \bibinfo{author}{Jacques \surnamestart Fleuriot\surnameend}
  (\bibinfo{year}{2010}): \emph{\bibinfo{title}{Automation for {{Geometry}} in
  {{Isabelle}}/{{HOL}}.}}
\newblock In: {\sl \bibinfo{booktitle}{{{PAAR}}@ {{IJCAR}}}}, pp.
  \bibinfo{pages}{84--94}.

\bibitemdeclare{inproceedings}{meikle2003}
\bibitem{meikle2003}
\bibinfo{author}{Laura~I. \surnamestart Meikle\surnameend} \&
  \bibinfo{author}{Jacques~D. \surnamestart Fleuriot\surnameend}
  (\bibinfo{year}{2003}): \emph{\bibinfo{title}{Formalizing {{Hilbert}}'s
  {{Grundlagen}} in {{Isabelle}}/{{Isar}}}}.
\newblock In \bibinfo{editor}{David \surnamestart Basin\surnameend} \&
  \bibinfo{editor}{Burkhart \surnamestart Wolff\surnameend}, editors: {\sl
  \bibinfo{booktitle}{Theorem {{Proving}} in {{Higher Order Logics}}}},
  \bibinfo{series}{Lecture {{Notes}} in {{Computer Science}}},
  \bibinfo{publisher}{{Springer}}, \bibinfo{address}{{Berlin, Heidelberg}}, pp.
  \bibinfo{pages}{319--334}, \doi{10.1007/10930755\_21}.

\bibitemdeclare{article}{minkowski1908}
\bibitem{minkowski1908}
\bibinfo{author}{Herrman \surnamestart Minkowski\surnameend}
  (\bibinfo{year}{1908}): \emph{\bibinfo{title}{Die {{Grundgleichungen}} F\"ur
  Die Elektromagnetischen {{Vorg\"ange}} in Bewegten {{K\"orpern}}}}.
\newblock {\sl \bibinfo{journal}{Nachrichten von der Gesellschaft der
  Wissenschaften zu G\"ottingen, Mathematisch-Physikalische Klasse}}, pp.
  \bibinfo{pages}{53--111}.

\bibitemdeclare{article}{mundy1986}
\bibitem{mundy1986}
\bibinfo{author}{Brent \surnamestart Mundy\surnameend} (\bibinfo{year}{1986}):
  \emph{\bibinfo{title}{Optical {{Axiomatization}} of {{Minkowski
  Space}}-{{Time Geometry}}}}.
\newblock {\sl \bibinfo{journal}{Philosophy of Science}}
  \bibinfo{volume}{53}(\bibinfo{number}{1}), pp. \bibinfo{pages}{1--30},
  \doi{10.1086/289289}.

\bibitemdeclare{article}{mundy1986a}
\bibitem{mundy1986a}
\bibinfo{author}{Brent \surnamestart Mundy\surnameend} (\bibinfo{year}{1986}):
  \emph{\bibinfo{title}{The {{Physical Content}} of {{Minkowski Geometry}}}}.
\newblock {\sl \bibinfo{journal}{The British Journal for the Philosophy of
  Science}} \bibinfo{volume}{37}(\bibinfo{number}{1}), pp.
  \bibinfo{pages}{25--54}, \doi{10.1093/oxfordjournals.bjps/37.1.25}.

\bibitemdeclare{inbook}{narboux2018}
\bibitem{narboux2018}
\bibinfo{author}{Julien \surnamestart Narboux\surnameend},
  \bibinfo{author}{Predrag \surnamestart Janicic\surnameend} \&
  \bibinfo{author}{Jacques \surnamestart Fleuriot\surnameend}
  (\bibinfo{year}{2018}): \emph{\bibinfo{title}{Computer-Assisted Theorem
  Proving in Synthetic Geometry}}, \bibinfo{edition}{1st} edition, pp.
  \bibinfo{pages}{21--60}.
\newblock \bibinfo{publisher}{Chapman and Hall/CRC}.

\bibitemdeclare{inproceedings}{palmer2018}
\bibitem{palmer2018}
\bibinfo{author}{Jake \surnamestart Palmer\surnameend} \&
  \bibinfo{author}{Jacques~D \surnamestart Fleuriot\surnameend}
  (\bibinfo{year}{2018}): \emph{\bibinfo{title}{Mechanising an {{Independent
  Axiom System}} for {{Minkowski Space}}-Time}}.
\newblock In: {\sl \bibinfo{booktitle}{Proceedings of the 12th {{International
  Conference}} on {{Automated Deduction}} in {{Geometry}}}}, pp.
  \bibinfo{pages}{64--79}.

\bibitemdeclare{book}{robb1936}
\bibitem{robb1936}
\bibinfo{author}{Alfred~A. \surnamestart Robb\surnameend}
  (\bibinfo{year}{1936}): \emph{\bibinfo{title}{Geometry of {{Time}} and
  {{Space}}}}.
\newblock \bibinfo{publisher}{{Cambridge University Press}}.

\bibitemdeclare{article}{schmoetten2021}
\bibitem{schmoetten2021}
\bibinfo{author}{Richard \surnamestart Schmoetten\surnameend},
  \bibinfo{author}{Jake~E. \surnamestart Palmer\surnameend} \&
  \bibinfo{author}{Jacques~D. \surnamestart Fleuriot\surnameend}
  (\bibinfo{year}{2021}): \emph{\bibinfo{title}{Towards {{Formalising Schutz}}'
  {{Axioms}} for {{Minkowski Spacetime}} in {{Isabelle}}/{{HOL}}}}.
\newblock {\sl \bibinfo{journal}{arXiv:2108.10868 [gr-qc]}}.
\newblock \eprint{2108.10868}.

\bibitemdeclare{book}{schutz1973}
\bibitem{schutz1973}
\bibinfo{author}{John~W. \surnamestart Schutz\surnameend}
  (\bibinfo{year}{1973}): \emph{\bibinfo{title}{Foundations of {{Special
  Relativity}}: {{Kinematic Axioms}} for {{Minkowski Space}}-{{Time}}}}.
\newblock {\sl \bibinfo{series}{Lecture {{Notes}} in {{Mathematics}}}}
  \bibinfo{volume}{361}, \bibinfo{publisher}{{Springer Berlin Heidelberg}},
  \bibinfo{address}{{Berlin, Heidelberg}}, \doi{10.1007/BFb0066798}.

\bibitemdeclare{article}{schutz1981}
\bibitem{schutz1981}
\bibinfo{author}{John~W. \surnamestart Schutz\surnameend}
  (\bibinfo{year}{1981}): \emph{\bibinfo{title}{An Axiomatic System for
  {{Minkowski}} Space\textendash Time}}.
\newblock {\sl \bibinfo{journal}{Journal of Mathematical Physics}}
  \bibinfo{volume}{22}(\bibinfo{number}{2}), pp. \bibinfo{pages}{293--302},
  \doi{10.1063/1.524877}.

\bibitemdeclare{book}{schutz1997}
\bibitem{schutz1997}
\bibinfo{author}{John~W. \surnamestart Schutz\surnameend}
  (\bibinfo{year}{1997}): \emph{\bibinfo{title}{Independent {{Axioms}} for
  {{Minkowski Space}}-{{Time}}}}.
\newblock \bibinfo{publisher}{{CRC Press}}.

\bibitemdeclare{phdthesis}{scott2008}
\bibitem{scott2008}
\bibinfo{author}{Phil \surnamestart Scott\surnameend} (\bibinfo{year}{2008}):
  \emph{\bibinfo{title}{Mechanising {{Hilbert}}'s {{Foundations}} of
  {{Geometry}} in {{Isabelle}}}}.
\newblock \bibinfo{type}{Master's thesis}, \bibinfo{school}{School of
  Informatics}, \bibinfo{address}{{The University of Edinburgh}}.

\bibitemdeclare{phdthesis}{scott2015}
\bibitem{scott2015}
\bibinfo{author}{Phil \surnamestart Scott\surnameend} (\bibinfo{year}{2015}):
  \emph{\bibinfo{title}{Ordered Geometry in {{Hilbert}}'s {{Grundlagen}} Der
  {{Geometrie}}}}.
\newblock \bibinfo{type}{{{PhD Thesis}}}, \bibinfo{school}{The University of
  Edinburgh}, \bibinfo{address}{{School of Informatics}}.

\bibitemdeclare{inproceedings}{scott2011}
\bibitem{scott2011}
\bibinfo{author}{Phil \surnamestart Scott\surnameend} \&
  \bibinfo{author}{Jacques \surnamestart Fleuriot\surnameend}
  (\bibinfo{year}{2011}): \emph{\bibinfo{title}{An {{Investigation}} of
  {{Hilbert}}'s {{Implicit Reasoning}} through {{Proof Discovery}} in
  {{Idle}}-{{Time}}}}.
\newblock In \bibinfo{editor}{Pascal \surnamestart Schreck\surnameend},
  \bibinfo{editor}{Julien \surnamestart Narboux\surnameend} \&
  \bibinfo{editor}{J{\"u}rgen \surnamestart {Richter-Gebert}\surnameend},
  editors: {\sl \bibinfo{booktitle}{Automated {{Deduction}} in {{Geometry}}}},
  \bibinfo{series}{Lecture {{Notes}} in {{Computer Science}}},
  \bibinfo{publisher}{{Springer}}, \bibinfo{address}{{Berlin, Heidelberg}}, pp.
  \bibinfo{pages}{182--200}, \doi{10.1007/978-3-642-25070-5\_11}.

\bibitemdeclare{article}{stannett2014}
\bibitem{stannett2014}
\bibinfo{author}{Mike \surnamestart Stannett\surnameend} \&
  \bibinfo{author}{Istv{\'a}n \surnamestart N{\'e}meti\surnameend}
  (\bibinfo{year}{2014}): \emph{\bibinfo{title}{Using {{Isabelle}}/{{HOL}} to
  {{Verify First}}-{{Order Relativity Theory}}}}.
\newblock {\sl \bibinfo{journal}{Journal of Automated Reasoning}}
  \bibinfo{volume}{52}(\bibinfo{number}{4}), pp. \bibinfo{pages}{361--378},
  \doi{10.1007/s10817-013-9292-7}.

\bibitemdeclare{article}{szekeres1968}
\bibitem{szekeres1968}
\bibinfo{author}{G.~\surnamestart Szekeres\surnameend} (\bibinfo{year}{1968}):
  \emph{\bibinfo{title}{Kinematic Geometry; an Axiomatic System for
  {{Minkowski}} Space-Time: {{M}}. {{L}}. {{Urquhart}} in {{Memoriam}}}}.
\newblock {\sl \bibinfo{journal}{Journal of the Australian Mathematical
  Society}} \bibinfo{volume}{8}(\bibinfo{number}{2}), pp.
  \bibinfo{pages}{134--160}, \doi{10.1017/S1446788700005188}.

\bibitemdeclare{incollection}{walker1959}
\bibitem{walker1959}
\bibinfo{author}{A.~G. \surnamestart Walker\surnameend} (\bibinfo{year}{1959}):
  \emph{\bibinfo{title}{Axioms for {{Cosmology}}}}.
\newblock In \bibinfo{editor}{Leon \surnamestart Henkin\surnameend},
  \bibinfo{editor}{Patrick \surnamestart Suppes\surnameend} \&
  \bibinfo{editor}{Alfred \surnamestart Tarski\surnameend}, editors: {\sl
  \bibinfo{booktitle}{Studies in {{Logic}} and the {{Foundations}} of
  {{Mathematics}}}}, {\sl \bibinfo{series}{The {{Axiomatic
  Method}}}}~\bibinfo{volume}{27}, \bibinfo{publisher}{{Elsevier}}, pp.
  \bibinfo{pages}{308--321}, \doi{10.1016/S0049-237X(09)70036-9}.

\bibitemdeclare{techreport}{wiedijk2000}
\bibitem{wiedijk2000}
\bibinfo{author}{Freek \surnamestart Wiedijk\surnameend}
  (\bibinfo{year}{2000}): \emph{\bibinfo{title}{The {{De Bruijn}} Factor}}.
\newblock \bibinfo{type}{Technical Report}, \bibinfo{institution}{{Department
  of Computer Science}}, \bibinfo{address}{{Nijmegen University}}.

\end{thebibliography}
\end{document}